\input harvmac

\def\l{\Lambda}
\def\su{SU( N - 1 )}
\def\sp{SP ({N \over 2})}
\def\sun{SU(N)}
\def\p{\partial}
\def\abs#1{\left | #1 \right|}


\def\drawbox#1#2{\hrule height#2pt 
        \hbox{\vrule width#2pt height#1pt \kern#1pt 
               \vrule width#2pt}
               \hrule height#2pt}


\def\Fund#1#2{\vcenter{\vbox{\drawbox{#1}{#2}}}}
\def\Asym#1#2{\vcenter{\vbox{\drawbox{#1}{#2}
              \kern-#2pt       
              \drawbox{#1}{#2}}}}
 
\def\Yfund{\Fund{6.5}{0.4}}
\def\Yasymm{\Asym{6.5}{0.4}}

\Title{hep-th/9608147, SCIPP 96/35}
{\vbox{\centerline{Dynamical Supersymmetry Breaking}
\centerline{versus Run-away Behavior  }
\centerline{in Supersymmetric Gauge Theories}}}
\bigskip
\centerline{
Yuri Shirman\footnote{$^\ast$}{yuri@scipp.ucsc.edu}}
\bigskip
{\vbox{\centerline{\it Santa Cruz Institute for Particle Physics}
\centerline{\it University of California, Santa Cruz,
CA 95064}}}
\bigskip
\bigskip
\baselineskip 18pt
\centerline{{\bf Abstract}}
\noindent
We consider Dynamical Supersymmetry Breaking (DSB)
in models with classical flat directions. We analyze
a number of examples, and develop a 
systematic approach to determine if classical flat 
directions are stabilized in the full quantum theory, or
lead to run-away behavior. In some cases pseudo-flat
directions remain even at the quantum level before
taking into account corrections to the K\"ahler potential.
We show that in certain limits these corrections are
calculable. In particular, we find that in the Intriligator-Thomas 
$SU(2)$ and its generalizations, a 
potential for moduli is generated. Moreover, there is a 
region of the parameter space where K\"ahler potential
corrections lead to  calculable (local) minima
at large but finite distance from the origin.

\Date{8/96}


\def\np#1#2#3{Nucl. Phys. B{#1} (#2) #3}
\def\pl#1#2#3{Phys. Lett. {#1}B (#2) #3}

\def\physrev#1#2#3{Phys. Rev. {D#1} (#2) #3}

\def\th#1{hep-th/#1}
\def\ph#1{hep-ph/#1}
\def\ptp#1#2#3{Prog. Theor. Phys. #1 (#2) #3}

\nref\ads{I. Affleck, M. Dine, and N. Seiberg, \np{256}{1985}{557}.}
\nref\newdsb{K. Intriligator, S. Seiberg, S. Shenker,
\pl{342}{1995}{152}, \ph{94102203};
M. Dine, A. Nelson, Y. Nir, Y. Shirman,
\physrev{53}{1996}{2658}, \th{9507378}; 
E. Poppitz, S. Trivedi,
\pl{365}{1996}{125}, \th{9507169};
P. Pouliot, \pl{367}{1996}{151}, \th{9510148};
P. Pouliot, M. Strassler, \pl{375}{1996}{175},
\th{9602031};
A. Nelson, \pl{369}{1996}{277}, \th{9511350};
C.-L. Chou, \th{9605108}; 
C. Csaki, W. Skiba, M. Schmaltz, \th{9607210}; 
K. Intriligator, S. Thomas, \th{9608046}.}
\nref\IT{K. Intriligator, S. Thomas, \th{9603158}.}
\nref\izawa{K. Izawa, T. Yanagida, \ptp{95}{1996}{829},
\th{9602180}.}
\nref\randall{C. Csaki, L. Randall, W. Skiba,
\th{9605108}.}
\nref\randalleigh{C. Csaki, L. Randall, W. Skiba,
R. Leigh, \th{9607021}.}
\nref\pst{E. Poppitz, Y. Shadmi, S. Trivedi, \th{9606184}.}
\nref\psttwo{E. Poppitz, Y. Shadmi, S. Trivedi, \th{9605113}.}

\newsec{Introduction}
If supersymmetry is relevant to the real world it is
important to investigate mechanisms by which it could be
broken. Models of dynamical supersymmetry breaking
(DSB) have been known for some time \ads. Recently,
many more models have been constructed \refs{\newdsb - 
\psttwo}.
Traditionally, one of the requirements in the construction
of the models with DSB was that they not possess
classical flat directions, and many more examples of this
kind have been studied recently. This requirement
was due to the observation \ads\ that when classical
flat directions are lifted by non-perturbative dynamics one 
usually finds run-away behavior.

It was pointed out  in \IT\
that this may not be the case if part of the gauge group
remains unbroken  with coupling which does not
tend to zero along the classical flat direction.
An example with $SU(2)$ gauge
group and matter transforming as 4 $SU(2)$ doublets
and 6 gauge singlets
(as well as generalizations for $SU(N)$ and $SP(N)$
groups) was constructed in
\refs{\IT, \izawa} where effects of the strong dynamics led
to non-vanishing energy along the classical flat direction.
However, a pseudo-flat direction remained even in the 
presence of the SUSY breaking superpotential.
The location of the minimum along this direction is determined
by the quantum corrections to the K\"ahler potential.
Due to the strongly coupled nature 
of the theory, these corrections could not
be determined, leaving open the possibility that the 
minimum is infinitely
far along the flat direction. Somewhat similar models
based on dual gauge groups were constructed in
\refs{\IT, \randall, \randalleigh}.

In this note we will suggest a systematic approach
to the analysis of the physics along classical flat
directions. We will discuss several examples, and
identify properties of the models which lead to
(non) stabilization of the classical flat directions.
In fact it is easy to summarize our approach to the
question of the possible run-away behavior.
Consider a model with classical flat directions
(assume for simplicity that there is a single direction
described by
modulus $S$). Can quantum effects stabilize the moduli near
the origin, or there will be a run-away behavior and
no stable vacuum? To answer this question
let us consider the theory at large vacuum
expectation value (vev) of $S$. It is convenient to
split the scalar potential into two parts\foot{This can be
done either in the full theory or in the low energy
effective theory applicable for large $S$.}:
\eqn\gensplit{V = V_L + V_S,}
where $V_L$ is the scalar potential of the ``reduced"
theory with $S$ considered as a parameter ($V_L$ may or may not
depend
on $S$), and $V_S = \abs{\p W
\over \p S}^2$. One should
first ask whether there are solutions to the equation
\eqn\genlow{V_L = 0.}
If there are no solutions to this equation, then the
energy for fixed moduli is proportional to the scale
characterizing the ``reduced" theory,
$V \sim \Lambda_L^4$.
If  $\Lambda_L$ grows with $S$, then there is an
effective potential for the modulus, and no run-away
behavior. This does not yet mean that supersymmetry is
broken in the full theory -- to establish that one needs
to analyze the theory near the origin of the moduli space. We
will describe an example of such a model in 
\S 5.
Another logical possibility is that $\Lambda_L
\rightarrow 0$ when $S \rightarrow \infty$.
If the same is true for $V_S$, then there is a run-away
direction in the full theory.

In the following sections we will be mostly considering
models such that \genlow\ {\it will} have a solution for every
value of $S$. In such models the question of run-away
behavior will become a question about the behavior
of $V_S$ for large $S$ (subject to \genlow). We will
discuss examples in which  run-away directions
persist in the full theory as well as examples where
classical flat directions are stabilized sufficiently far
from the origin. We will also consider an example where
moduli are stabilized near the origin, and analysis in the region
of the strong coupling is required to establish SUSY breaking.

\newsec {SUSY QCD with singlets}
We will begin by considering massive 
SUSY QCD with $N_f=N_c$.
It is known that in this model the classical moduli
space is modified quantum mechanically
\ref\seiberg{N. Seiberg, \physrev{49}{94}{6857},
\th{9402044};
K. Intriligator, N. Seiberg \th{9509066}.}:
\eqn\modspace{ det (M) -B \bar B = \Lambda^{2N},}
where $M_{ij}=Q_i \bar Q_j$, $B = Q^N$,
$\bar B = \bar Q ^N$. This constraint can be enforced
by a Lagrange multiplier term, so that the full
superpotential is:
\eqn\qcdw{W = A (det M - B \bar B - \Lambda^{2N})
+ m_{ij} M_{ij}.}
In the vacuum mesons have the following vev's:
\eqn\qcdvacuum{ M_{ij} = ( det (m) \ \Lambda^{2N})
^{1 \over N} \left({1 \over m}\right)_{ij}^{-1}.}

It was found in \refs{\IT , \izawa}
that a simple
modification of this model breaks supersymmetry
dynamically. Instead of giving mass to the quarks
one can couple them to $N^2 + 2$
gauge singlets:
\eqn\qcdtree{W_{tree} = \sum_{ij}^N
\lambda_{ij} S_{ij} M_{ij} + X_1 B +
X_2 \bar B.}
This superpotential lifts all flat directions associated
with SU(N), but classically there are still flat
directions for the singlets.
Quantum mechanically the 
energy is non-vanishing along these flat directions \IT.
When the singlets have large vev's quarks become massive and
the low energy
theory is pure Yang-Mills with the scale
\eqn\pureym{\Lambda_{YM}^{3N} = det ( \lambda S )
\Lambda^{2N}.}
Gaugino condensation gives rise to a superpotential
\eqn\ymsuper{W = \Lambda_{YM}^3 = ( det ( \lambda S )
\Lambda^{2N})^{1 \over N} \sim \lambda S \Lambda^2,}
and the energy does not vanish in the limit
$S \rightarrow \infty$.
Classically there is a supersymmetric minimum
at the origin of the $\sun$ moduli space. This, however,
is not compatible \refs{\IT , \izawa} with the quantum constraint 
\modspace.
To see this let us
split scalar potential into two terms
\eqn\qcdscalar{V = V_L + V_S =
\left( \abs {\p W \over \p A}^2
+ \sum_{ij} \abs
{\p W \over \p M_{ij}}^2 \right)
+ \sum_{ij}
\abs {\p W \over \p S_{ij}}^2 }
Here the $V_L$ is just the scalar potential of SQCD
with singlet vev's playing the role of masses. It
vanishes whenever \qcdvacuum\ is satisfied.
The second term, however, contains a sum over all
fields involved in the quantum constraint \modspace, and
therefore can not vanish.

The potential of eqn. \qcdscalar\ is
independent of $S$. Therefore,
at the level of
the analysis performed so far there is still an exact
flat direction even though supersymmetry is broken.
In \IT\  various possibilities
for the behavior of the theory at large $S$ were
considered, including the existence of the minimum
for $S \rightarrow \infty$. However, one can easily see
that this is not the
case. We will consider $SU(2)$ with 4 doublets and
6 singlets as an example. For simplicity, take the 
tree level
superpotential to possess global $SU(4)$ symmetry:
\eqn\globalsufour{W_{tree} = \lambda \sum_{i < j}^4
S_{ij} Q_i Q_j.}
Note that the energy is
proportional to the coupling constant squared
$\abs{\lambda}^2$. $\lambda$ should be interpreted as
the running coupling constant, $\lambda(S)$. (More
precisely, it is $\lambda(\lambda S)$. This can be 
determined by, for example, carefully implementing
scale matching prescription \pureym\ so that analyticity
of the gauge coupling function is preserved at the two 
loop level.)
The $S$ dependence of the coupling constant corresponds to
the renormalization of the K\"ahler potential. For large $S$
and small $\lambda$ a perturbative calculation is valid
and gives
\eqn\betafunction{\beta_\lambda = {8 \over 16 \pi^2}
\lambda^3 -
 {3 \over 16 \pi^2} \lambda g^2.}
If $\lambda$ is taken sufficiently large, the right-hand
side of \betafunction\ is always positive, and the coupling
constant $\lambda$ is not asymptotically free. This corresponds
to a growing potential for $S$, and the minimum is necessarily
near the origin\foot{This description breaks down at large
scales as
$\lambda$ hits its Landau singularity. At those scales an 
underlying
microscopic description of the theory should be used. However,
there
still exists (at least a local) minimum near the origin.}
(its exact location can not be determined by
this method). A somewhat more interesting situation arises if
we choose the coupling constant so that at a certain scale
$M \gg \l$ the right-hand side of \betafunction\ vanishes.
Note that ${\p (\lambda / g) \over \p t} > 0$
in this model, and there is no fixed point for the Yukawa coupling
constant. 
As a result $\lambda$ exhibits an asymptotically free behavior
below $M$, and grows above $M$. In such a regime scalar
the potential of the model has a minimum at $S \sim {M \over \lambda}$.

Now, let us consider one more modification
of the model, namely, suppose that only $N - 1$ flavors
are massive while the last flavor
couples to the gauge singlet
\eqn\onesinglet{W_{tree} = \sum_{ij}^{N-1}
m_{ij} M_{ij} + S M_{NN}.}
Splitting the scalar potential into two terms in analogy
with \qcdscalar\ we see that $V_S = \vert M_{NN}
\vert ^2$. This can be infinitesimally small with
some other mesons having infinitely large vev's in
such a way that both $V_L = 0$, and the quantum
constraint \modspace\ is satisfied. From \qcdvacuum\
we find\foot{Substituting $m_{NN} =\ < S >$.}:
\eqn\qcdscaling{ M_{ij} \sim S^{1 \over N}, \ \
M_{NN} \sim S^{ - {N-1 \over N}},}
where $i,j = 1\ldots N-1$. We see that SUSY is restored
when $S \rightarrow \infty$. One can reach the same
conclusion in any modification of the
$\sun$ theory with $N$ flavors where some of the classical
flat directions are lifted by the mass terms rather than
by coupling to gauge singlet fields.

Before going on to the next section, where we
will use these results, let us briefly mention
analogous results for the model based on $SP(N)$
gauge group with $N + 1$ flavors in the fundamental
representation. The quantum modified constraint is given
by \ref\intpouliot{K. Intriligator, P. Pouliot,
\pl{353}{1995}{471}, \th{9506006}.}:
\eqn\spmodspace{ Pf ( M ) = \Lambda ^{2 (N+1)},}
where $M_{ij} = Q_i Q_j$. In the massive case
the vacuum is given by:
\eqn\spvacuum{M_{ij} = ( Pf (m) \ \Lambda^{2 (N + 1)})
^{1 \over N + 1} \left( {1 \over m}\right)_{ij}^{- 1}}

When quarks are coupled to $(N + 1)^2$ gauge singlets,
the scalar potential is given by \qcdscalar\ with $V_L$
being the potential of the massive $SP(N)$
theory with $N + 1$ flavors. $V_S = 0$ is not
compatible with the quantum constraint and
SUSY is broken \refs{\IT , \izawa}. When $N$ flavors are
given mass, while
the last flavor is coupled to a gauge singlet, S,
the meson vev's scale as:
\eqn\spscaling{ M_{ij} \sim S ^{1 \over N+1},\ \
\ M_{2N+1,2N+2} \sim  S ^ {- {N \over N+1}},}
where $i,j = 1 \ldots 2N$, and the
energy vanishes as $S \rightarrow \infty$.

\newsec{ ${\bf SU(N - 1)  \times SP ({N \over 2})}$ versus
${\bf SU(N - 1)}  \times {\bf SU(N)}$ model}

Our next example will be a model described
in \IT\ and based on $\su \times \sp$
gauge group\foot{Note that we
shifted $N$ relative to definition
in \IT, and $N$ is even.}
with matter transforming as $Q\ (N-1, N)$,
$L\ (1, N)$, $\bar Q_i\ (\overline {N-1}, 1)$,
$i = 1\ldots N$.
The tree level superpotential of this model is given by:
\eqn\sptree{W_{tree} = \lambda Q L \bar Q_2 +
{1 \over M}
\sum_{i,j>2}^{N} \lambda_{ij} Q^2 \bar Q_i \bar Q_j.}
This superpotential leaves classical flat directions
associated with the $SU(N - 1)$ antibaryons
$\bar B = \bar Q^{N - 1} = v^{N - 1}$
(we will denote the antiquark vev by $v$).
There is also a non-perturbative
superpotential\foot{We will denote $\l_{\su}$ as
$\l_1$, and both $\l_{\sp}$ and later
$\l_{\sun}$ as $\l_2$. }
\eqn\spdyn{W_{np} = {det \hat q - \bar B \hat q B \over
\l_1^{2N-3}} + A ( Pf( M ) - \l_1^{2N-3} \l_2^{N+3}),}
where
$\hat q_i = q_i = Q \bar Q_i$,
$q_{N+1} = L$, $q_{N + 2} = B = Q^{N-1}$,
$\bar B = \bar Q^{N-1}$, and
$M_{ij} = q_i q_j$.

First, consider the limit $\l_1 \gg \l_2$.
$\su$ confines and below the scale $\l_1$
the effective theory is $\sp$ with $N_f = {N \over 2} +1$
flavors. The correct description of physics is given
in terms of the following canonically normalized fields
\eqn\splowfields{q_i = {Q \bar Q_i \over \l_1}, \
q_{N+1} = L,\
q_{N+2} = { B \over \l_1^{N-2}} = { Q^{N-1} \over
\l_1^{N-2}}, \
S = {\bar B
\over \l_1^{N-2}} ={ \bar Q^{N-1} \over \l_1^{N-2}}.}
The superpotential of equations \sptree\ and
\spdyn\ turns into
\eqn\spwlow{W = \sum_{ij > 1}^{N+1} m_{ij} q_i q_j +
{Pf^\prime ( M ) \over \l_1^{N-3}} - S_i q_i q_{N+2} +
\tilde A (Pf ( M ) - \tilde \l_2^{N+2}).}
Here $m_{2,N+1} = \lambda \l_1$,
$m_{ij} = {\lambda_{ij} \l_1^2 \over M}$,
$\tilde \l_2^{N+2} = {\l_2^{N+3} \over \l_1}$,
$M_{ij} = q_i q_j$, $\tilde A = \l_1^{2 N -2} A$,
and $Pf^\prime$ denotes the
Pfaffian over the first ${N \over 2}$ flavors.
Without loss of generality we can set
$S_i = 0$, $M_{i,N+2} = 0$ for $i \ne 1$ and be left
with the model
described in the
previous section (with the change $N \rightarrow
{N\over 2}$).
We conclude that there is a run-away direction
$S \rightarrow
\infty$ along which scalar potential goes to zero:
\eqn\sprunawaylow{V_S = \abs {\p W \over \p S}^2
\sim S^{-{2N \over N+2}}.}

This solution, however, is only valid for $S \ll
\Lambda_1$ \foot{We thank Scott Thomas for discussion
of this point.}. We have no tools available to analyze
the behavior of the model at scales $S \sim \Lambda_1$.
At  scales $S \gg \l_1$ the relevant degrees of freedom are
the elementary ones.
We still can make use of our previous results.
Namely, as long
as $V_L$
is {\it exactly} zero in terms of meson degrees of
freedom, it will be zero in terms of quark degrees of freedom.
Therefore, we only have to reconsider behavior of $V_S$:
\eqn\larges{V_S = \abs{ { \p W \over
\p S} {\p S \over \p \bar Q}
} ^2 =
(N - 1)^2 \abs{ M_{1,N+2}}^2 v^{2(N-2)}
\sim  v^{ 2 {N-4 \over N+2}}.}
We see that for 
$N > 4$ it increases along the classical flat direction.
Thus the classical flat direction is stabilized
quantum mechanically. Analysis of the theory in the finite
region of the field space \IT\ shows that SUSY is broken.
It is
interesting to note that neglecting the quantum modified
constraint in $\sp$ group (second term in \spdyn)
one can find supersymmetric minima with vev's of
the fields of order $\l_1$. However, these minima
are incompatible with the quantum modified constraint.

For $N = 4$ (that is for the $SU(3) \times SP(2)$
model) the potential is constant independent of moduli. 
The situation is somewhat
analogous to the example in \S 2. The $\beta$-function
for the renormalizable coupling in eqn. \sptree\ is
\eqn\dualbeta{\beta_\lambda = {8 \over 16 \pi^2}
\lambda^3 - {\lambda \over 16 \pi^2} 
({16 \over 3} g_1^2 + 5 g_2^2).}
If $\lambda$ is chosen so that right-hand side
of \dualbeta\ is positive, it grows with the scale.
In the limit $g_1 \gg g_2$ there is a fixed point
for the ratio $\lambda / g_1$. If $\lambda$ is chosen
sufficiently small it is bounded from above by
its fixed point value and is, therefore, asymptotically
free. 
Analysis of the non-renormalizable
terms is more complicated and requires at least some
assumptions about the properties of the underlying
microscopic theory. Nevertheless, the general 
conclusion is that corrections to the K\"ahler 
potential may or 
may not lead to run-away behavior depending 
on the choice of the parameters of the model.

We would like to compare these results to the
behavior of the model \pst\ based on $\su \times \sun$
gauge group with matter in the fundamental
representations:
$Q\ (N -1, N)$, $\bar L_a\ (1, N)$,
and $\bar Q_i\ (\overline {N - 1}, 1)$,
where $a=1\ldots N - 1$, and $I=1\ldots N$.
The tree level superpotential is given by:
\eqn\sutree{W_{tree} = \sum_{ia} \lambda_{ia} 
Q \bar L_a \bar Q_i.}
As in the previous case there are classical
flat directions parameterized by $\su$ antibaryons.
The non-perturbative superpotential is given by
\eqn\sudyn{W_{np} = {det ( q ) -  B q \bar B \over
\l_1^{2N-3}} +
A ( det (q \bar q) - det ( q ) \ det (\bar q ) -
{\l_1}^{2N-3} {\l_2}^{2N+1}),}
where $q = Q \bar Q$, $B = \bar q_N = Q^{N-1}$,
$\bar B = \bar Q^{N-1}$,
and, finally, $\bar q_a = L_a$.
Let us comment on the similarities of the two models.
Both of them can be constructed in the following way.
Start with $SU(N-1)$ gauge group and N flavors of
matter in the fundamental representation and gauge a
subgroup of the global symmetry (one also needs to add
matter transforming with quantum numbers of the second
gauge group to cancel anomalies). In both models
the $\su$ gauge group is confining, it also has a flat
direction parameterized by the antibaryons.
The second gauge group is one flavor short of developing
a quantum constraint \modspace\ or \spmodspace. Such a
constraint develops in the effective theory upon confinement
of $SU(N-1)$. Therefore we can expect a similar
behavior.

In fact this is true for
$S = {\bar B \over \l_1^{N-2}} \ll \l_1$.
As in the Intriligator - Thomas model, the $\su$ group
confines and the effective theory is $\sun$ with $N$
flavors. $N-1$ of them are massive, while
the last flavor is coupled to a $\sun$ singlet field.
Using \qcdscaling\ we find for $S \ll \l_1$
\eqn\surunawaylow{V_S = \abs{ \p W \over \p S }^2
\sim S^{- 2 {N-1 \over N}}}
in complete analogy with \sprunawaylow. On the other hand
above the $\su$ confinement scale result is qualitatively
different
\eqn\surunawayhigh{
V_S = (N - 1 )^2 \abs{ M_{NN} }^2 v^{2(N-2)}
\sim v^{- {2 \over N}}.}
There is still a run-away direction in the
full quantum theory\foot{It is possible to add
non-renormalizable operators to the superpotential
lifting remaining classical flat directions. In such
a case SUSY is broken \pst.}.

We saw that classical flat directions are stabilized
due to the special
group structure and matter content
of the Intriligator-Thomas model. It is
easy to repeat this analysis in the limit
$\l_1 \ll \l_2$. Then we would find
exactly the same results as before but
stabilization  of the classical flat direction
would appear to be a consequence of the
presence of the non-renormalizable terms
in the $\su \times \sp$ model.

\newsec{SU(N - M) $\times$ SU(N) models}
A simple generalization \psttwo\ of the $SU(N-1) \times
SU(N)$ model of the previous section is based on
$SU(N - M) \times SU(N)$ gauge group with 
the following matter content:
$Q (N-M, N)$, $\bar Q_i\ ( \overline {N-M}, 1 )$, and
$\bar L_a\ (1, \bar N)$, where $I = 1 \ldots N$,
$a = 1\ldots N-M$.
The tree level superpotential is given by:
\eqn\nminusm{W_{tree} =
\lambda_{ia} Q \bar L_a \bar Q_i.}
This lifts all flat directions except those
associated with the $SU(N - M)$ antibaryons
$\bar B^{i_1...i_M} = (\bar Q^{N-M})^{i_1...i_M}$.
Along the classical flat directions the $SU(N)$ gauge
group remains unbroken and its quantum effects
have to be taken into account.
$SU(N)$ gauge dynamics generates a non-perturbative
superpotential:
\eqn\nmwnp{W_{np} = \left ({\Lambda^{2N+M}
\over det ( Q \bar L )} \right)^{1 \over M}.}
This results in a scalar potential which can be written
as
\eqn\nmscalar{V = V_L + V_{\bar Q},}
where $V_L$ is the scalar potential of massive SQCD
with $N - M$ flavors and $\bar Q$ vev's acting as masses,
while
\eqn\nmbarq{V_{\bar Q} =  \abs{ \p W \over
\p \bar Q_i}^2.}
Again using \qcdvacuum\ we can easily see
that when $V_L = 0$
and $\bar B \sim \bar Q^{N-M} \rightarrow \infty$,
the second term in \nmscalar\ is
also vanishing: $V_{\bar Q} \rightarrow 0$.
SUSY, therefore, remains unbroken\foot{As in the
$\su \times \sun$ model it is possible to add
non-renormalizable terms to the superpotential
\nminusm. At least for $M=2$ all flat directions
can be lifted. Then for odd $N$ supersymmetry is
broken \psttwo.}.

It is possible to construct a dual for
the $SU(N - M) \times SU(N)$  model.
It is instructive to see
how SUSY is restored in the dual picture. The dual gauge
group is $SU(M) \times SU(N)$ with fields
transforming as
$q\ (M, \bar N)$, $\bar q_i\ (\bar M, 1)$,
$M_i\ (1, N)$, $\bar L_a\ (1, \bar N)$,
$I = 1...N$, $a = 1...N-M$.

The tree level superpotential consists of the terms
inherited from the electric theory as well as terms
added in the process of dualizing
\foot { The parameter $\mu$ appearing here relates scales
of electric and magnetic theories.}
\eqn\dualtree{W_{tree} = \lambda_{ai} \mu M_i \bar L_a
+ M_i q \bar q_i.}
This superpotential
leaves flat directions parameterized by the $SU(M)$ antibaryons
$\bar q^M \sim \bar Q^{N-M}$, which are exactly the flat
directions of the electric theory.
In the dual description
$SU(N)$ has $N$ flavors, and therefore, develops a quantum
constraint\foot{We will use here a notation
$q_{M+a} = \bar L_a$.}
\eqn\dualconstraint{W_{np} = A ( det (q M) -
b \bar b - \tilde \Lambda^{2N}),}
where $b = M^N$, $\bar b = q^N$, and $\tilde \l$ is
the scale of the dual theory.
Now we can use \qcdvacuum\ and \qcdscalar\ (remember
that $\bar q_i$ are $SU(N)$ singlets) to conclude
that the run-away behavior persists in the quantum theory,
and that the vacuum energy vanishes as $ \bar B
\rightarrow \infty$. This conclusion remains valid above
$\tilde \l$.

One clarification is needed here. The above discussion assumes
that at least in one description model is weakly coupled
and the K\"ahler potential is nearly canonical along the classical flat
direction. This is not always true. Suppose that we chose $M$ so that
both original $SU(N-M)$ and its dual $SU(M)$ groups are 
asymptotically free. Neglecting the $SU(N)$ dynamics the model has a 
fixed point where both descriptions are strongly coupled.
However, the scaling of the K\"ahler potential at the fixed point is
known, thus strong coupling effects can be taking into account,
and one finds that model still exhibits run-away behavior\foot{We
thank M. Peskin for the discussion of this point}.

\newsec{Some Other Models}

In this section we will briefly discuss models
for which a solution to $V_L = 0$ does not exist.
As an example we will use the $SU(N) \times SU(4)
\times U(1)$ model of ref. \randalleigh\
with the field content given by
$$A (\Yasymm, 1)_8,\
\ a (1, \Yasymm )_{-2N}, \ \ T (\Yfund , \Yfund )_{4-N},\ \
\bar F_i (\overline{ \Yfund} , 1)_{-4}, 
\ \ \bar Q_i (1, \overline {\Yfund} )_N,$$
where $i = 1 \ldots N$.
All classical flat directions of this model are
lifted if the tree-level superpotential is chosen to be
\eqn\randalltree{W_{tree} = A \bar F_1 \bar F_2 + 
A \bar F_3 \bar F_4 \ldots
+ A \bar F_{N-2} \bar F_{N-1}  + 
a \bar Q_2 \bar Q_3 +  \atop \hfill a \bar Q_4 \bar Q_4 + 
\ldots + a \bar Q_{N-1} \bar Q_1
+  T \bar F_1 \bar Q_1 + \ldots + T \bar F_N \bar Q_N.}
We will follow \randalleigh\ and set coefficients
of the terms $a \bar Q_i \bar Q_j$ to zero. Then the
model possesses classical flat directions. Along
the most general flat direction the $SU(4)$ gauge group is
completely broken while the $SU(N)$ gauge group remains
unbroken. Three flavors of the $SU(N)$ fields
become massive and the low energy effective theory has
$SU(N)$ gauge group with antisymmetric tensor,
$N-4$ antifundamentals, and tree-level superpotential
which raises all the flat directions of the ``reduced"
theory. Such a model is known to break
supersymmetry\foot{In general one has to
investigate in this way
all possible flat directions. In this case
if we chose to look in a more specialized flat
direction the low energy effective theory
would be different, but still supersymmetry breaking.}
with the vacuum energy
\eqn\liftedflat{V \sim \l_L^4 \sim
\left( v^4 \l^{2N-1} \right)^{4\over 2N+3},}
where $v$ denotes generic vev of the moduli.
We immediately see that a potential for the moduli is
generated, and there is no run-away behavior.
An analysis of the strong coupling dynamics near
the origin of the moduli space conducted in \randalleigh\
shows that supersymmetry is broken in the model.
Analogous conclusions apply to the $SU(N) \times SU(3) \times U(1)$
and $SU(N) \times SU(5) \times U(1)$ models of \refs{\randall,
\randalleigh}.

\newsec{Conclusions}
In this note we considered the behavior of models with classical
flat directions. We showed that it is convenient to analyze
physics along these directions in two stages.
First, one considers the moduli as fixed parameters. 
If the ``reduced''
theory breaks SUSY, and its scale increases with the moduli vev, 
then classical flat directions are stabilized. 
If the scale decreases then theory exhibits run-away
behavior.
If the ``reduced'' theory does not break SUSY
one has to consider dynamics for the moduli subject
to the requirement that potential of the ``reduced'' theory
vanishes. At different scales the theory may be best
described by different sets of variables. Therefore this 
analysis should be repeated at all relevant scales. 
The asymptotic behavior of the potential
can be determined by the analysis in the region where
moduli vev greatly exceed any dynamically generated scale of the
model. This means that the answer can usually be found 
most easily
from an analysis in 
terms of the elementary degrees of freedom.

In some cases one finds pseudo-flat direction even in the
presence of a non-perturbative superpotential. 
Then corrections to the K\"ahler potential determine the
location of the true vacuum. We showed that in certain regimes
such corrections are calculable and can be computed as 
renormalizations of the Yukawa couplings. 
Thus the potential at 
infinity is determined by asymptotic behavior of the coupling
constants of the model. In the Intriligator-Thomas
$SU(2)$ the non-asymptotically free nature of the Yukawa
coupling lead to the stabilization of the potential.
Moreover, we showed that for sufficiently small
coupling $\lambda$ 
there are (calculable) minima for large but finite 
values of the moduli fields. We also saw an example
where the potential could be destabilized due to
the existence of a quasi-fixed point
for the Yukawa coupling.

{\bf Acknowledgments.} We would like to thank 
Michael Dine, and Scott Thomas for useful 
discussions and reading of the manuscript. We would
also like to thank Michael Peskin for useful discussions. 
This work was supported in part by the 
U. S. Department of Energy.

\listrefs
\end